\documentclass[
  ,draft            
  ]
  {aipproc}

\layoutstyle{6x9}

\newcommand{\NP}[1]{{ Nucl.\ Phys.\ } {\bf  #1}}
\newcommand{\ZP}[1]{{ Z.\ Phys.\ } {\bf  #1}}

\newcommand{\PL}[1]{{ Phys.\ Lett.\ } {\bf  #1}}

\newcommand{\PR}[1]{{ Phys.\ Rev.\ } {\bf  #1}}
\newcommand{\PRL}[1]{{ Phys.\ Rev.\ Lett.\ } {\bf  #1}}

\newcommand{\lsim}{\raise.3ex\hbox{$<$\kern-.75em\lower1ex\hbox{$\sim$}}}
\newcommand{\ima}{{\mbox{Im}\,}}
\newcommand{\rea}{{\mbox{Re}\,}}
\newcommand{\be}{\begin{equation}}
\newcommand{\ee}{\end{equation}}

\begin{document}

\title{Light meson resonances from unitarized Chiral Perturbation Theory}

\author{J.R.Pel\'aez}{address={Dip. di Fisica. Universita' degli Studi and INFN,
 Firenze,
  Italy},
  altaddress={Departamento de F\'{\i}sica Te\'orica II\\
  Universidad Complutense\\28040 Madrid, Spain}
}

\author{A.G\'omez Nicola}{
  address={Departamento de F\'{\i}sica Te\'orica II\\
  Universidad Complutense\\28040 Madrid, Spain}
}

\begin{abstract}
 We report on our recent progress in the generation of resonant behavior
in unitarized meson-meson scattering amplitudes obtained from 
Chiral Perturbation Theory.  These amplitudes provide simultaneously 
a remarkable description of the resonance region up to 1.2 GeV 
as well as the low energy region,
since they respect the chiral symmetry expansion.
By studying the position of the poles in these amplitudes
it is possible to determine the mass and width of the associated resonances,
as well as to get a hint on possible classification schemes, 
that could be of interest for the spectroscopy of the scalar sector.
 \end{abstract}

\maketitle

\section{The light meson puzzle}

In this work we review our recent progress in determining the 
position of the poles \cite{prep} that appear associated to resonant behavior
in meson-meson scattering amplitudes, obtained from unitarized
one-loop Chiral Perturbation Theory \cite{GomezNicola:2001as}.
This apparently formal interest is motivated
by the spectroscopy of light mesons, whose present status is somewhat 
controversial. Poles in the second Riemann sheet of partial wave
scattering amplitudes are of relevance
because when they are close to the real, physical values 
of the center of mass energy $\sqrt{s}$, we can neglect all other
terms in the partial wave and simply write
\begin{eqnarray}
t(s)=\frac{R_R}{s-s_{pole}}=
\frac{R_R}{s-(\hbox{Re}\sqrt{s_{pole}})^2-(\hbox{Im}\sqrt{s_{pole}})^2
-i\,2\,\hbox{Re}\sqrt{s_{pole}}\;\hbox{Im}\sqrt{s_{pole}}}
\end{eqnarray}
where $R_R$ would be some real residue that can be calculated 
but is irrelevant for us here.
Furthermore, if by ``close to the real axis'' we mean
that Im$\sqrt{s_{pole}}\ll$Re$\sqrt{s_{pole}}$, then, we can approximate:
\begin{eqnarray}
t(s)\simeq
\frac{R_R}{s-(\hbox{Re}\sqrt{s_{pole}})^2
-i\,2\,\hbox{Re}\sqrt{s_{pole}}\;\hbox{Im}\sqrt{s_{pole}}}
\equiv
\frac{R_R}{s-M_R^2+i\,M_R\,\Gamma_R}
\label{BWform}
\end{eqnarray}
where in order to write our
equation in the familiar Breit-Wigner form, in the last step we have identified
$\sqrt{s_{pole}}\simeq M_R-i \Gamma_R/2$.
Breit-Wigner (BW) resonances yield the familiar and experimentally
distinct resonant shape in the cross section
and its associated fast phase movement, which increases by $\pi$
in a very small energy range. 
The quantum numbers of the resonances correspond
to those of the partial wave where the pole is sitting.

However, the farther away from the real axis the
poles are, the lousier becomes the connection with resonance parameters.
Let us remark that in order to have a  BW shape,
it is essential for the pole to be near the real axis, or
more quantitatively $M_R\gg\Gamma_R$. 
This allows us to neglect all other terms in the amplitude
as well as terms of order $\Gamma_R^2/M_R^2$.
Intuitively, the familiar resonances that are clearly seen or detected
are quasi bound states whose decay time
is large (their width is small) compared with their rest energy (their mass).
Of course, between a nice BW resonant shape and the continuum,
one could think of all intermediate situations, which,
naively correspond to changing the pole position  
from the vicinity of the real axis to  have an infinite imaginary part.
In other words, starting
from narrow resonances
and moving the pole to $-i\,\infty$, we get  broader structures, 
and finally, the continuum.

In particular, broad resonant structures seem to occur
in the scalar channels in meson-meson
scattering, where in the last decade there has been a renewed interest
\cite{newsigma,Dobado:1992ha}
on the longstanding controversy about
the existence of a broad scalar-isoscalar resonance in the low energy region:
 the so called $\sigma$, or $f_0(600)$ 
in the latest version of the Particle Data Group (PDG) Review \cite{PDG}. Its experimental evidence 
only from  $\pi\pi$ scattering is rather confusing, since it definitely
does not display a Breit-Wigner shape, although many groups have been able
to identify an associated pole in the
amplitude, but deep in the complex plane 
A similar or even more confusing situation occurs in $\pi K$
scattering, where another pole, the $\kappa$, has been suggested by
many groups \cite{kappa,Oller:1997ng}, but again there is no trace of a BW shape
in the scattering. For an compilation of $\sigma$ and $\kappa$ poles 
see the nice overview in \cite{vanBeveren:2002vw}. 

Let us remark that meson-meson scattering data  \cite{pipidata} 
are hard to obtain.
As a matter of fact
the problem is that 
they have been extracted from reactions
like meson-$N\rightarrow$meson-meson-$N$, but with assumptions
like a factorization of the 
four meson amplitude, or that only one meson is exchanged and that
it is more or less on shell, etc... All these approximations
introduce large systematic errors.
There are, however, other sources of information on 
meson-meson interactions like, for instance, the very precise
determination of a combination of $\pi\pi$ phase shifts
from $K_{l4}$ decays \cite{Pislak:2001bf}. At higher energies the decays of even heavier particles
can be also used to study the previously
mentioned and other
scalar resonances like the $f_0(980)$ or the $a_0(980)$.
For instance, very recently,
 results from charm decays \cite{charm}, seem to find
both the $\sigma$ and $\kappa$ poles in reasonable agreement
with the groups mentioned above, but the controversy about their
existence still lingers on.

Meson spectroscopy aims at classifying the bound states of QCD and
at identifying their nature, that is, what are they made of. 
Starting with the scalar-isoscalar sector, its relevance
 is twofold: First, one of the
most interesting features of QCD is its non-abelian nature, which 
implies that the carriers of the strong force, the gluons, interact among
themselves, contrary to what happens with photons in QED. A possible
consequence of this fact 
is the existence of bound states of gluons, or glueballs,
which will certainly be isoscalars. In particular,
the lightest ones are
expected to be also scalars. Naively, once all the members
of quark multiplets are identified in the scalar-isoscalar sector,
what remains, if any, are good candidates for glueballs. Of course,
the whole picture is much more messy due to mixing phenomena,
so that the resonances we actually see are a superposition of 
different kind of states. Second, it is also understood that
QCD has an spontaneous breaking of the chiral symmetry since its
vacuum is not invariant under chiral transformations. The study
of the scalar-isoscalar sector is relevant to understand the QCD vacuum,
which has precisely those quantum numbers.

Nevertheless, we should not forget the other channels, since we can
find there the other members of the multiplets, since
all the channels are related
by the chiral SU(3) symmetry of QCD. We cannot simply add BW resonances
to different channels without carefully taking into account this symmetry.
Concerning vector channels, there are clear
BW resonances like the $\rho(770)$ in $\pi\pi$ scattering
or the $K^*(892)$ in $\pi K$ scattering, that the meson spectroscopy
community identify with $q\bar{q}$ states. These are so clearly resonant
that ``vector meson dominance'' is basically enough to describe
the bulk of meson interactions.

\section{Poles from Chiral Symmetry and Unitarity}

The interest of our work in the context of
meson spectroscopy  is that we have been able 
to {\it generate} the resonant behavior present in meson-meson scattering.
Our amplitudes \cite{GomezNicola:2001as} have been obtained by unitarizing the one-loop
amplitudes obtained from Chiral Perturbation Theory 
(ChPT \cite{chpt}), which is
the most general effective Lagrangian built of pions, kaons and etas,
that respects the chiral symmetry constraints of QCD. 
However, since the 
ChPT amplitudes behave as polynomials at high energy, they violate
partial wave unitarity, which is imposed with unitarization methods:
in our case, the Inverse Amplitude Method (IAM) 
\cite{Dobado:1996ps,Dobado:1992ha}. Note that {\it the resonances are not included explicitly}.

Part of this program had been first been carried out for 
partial waves in the elastic region
\cite{Dobado:1996ps,Dobado:1992ha}, for which a simple single channel
approach could be used, finding the $\rho$ and $\sigma$ poles
in $\pi\pi$ scattering and that of  $K^*$ in $\pi K\rightarrow\pi K$. 
For coupled channel processes, 
an {\it approximate} form of this approach had already been shown
\cite{Oller:1997ng} to yield a remarkable description of 
the whole meson-meson
scattering data up to 1.2 GeV. When these partial waves were continued to
the second Riemann sheet,  several poles were found,
corresponding to the $\rho$, $K^*$, $f_0$, $a_0$, $\sigma$ and $\kappa$
resonances ( note that the $\kappa$ pole could have also been
obtained in the elastic single channel formalism ).
 The approximations were needed because at that time
not all the ChPT meson-meson amplitudes were known to one-loop.
Hence, in \cite{Oller:1997ng} 
only the leading order and the dominant s-channel loops were 
considered in the  calculation, neglecting crossed and tadpole loop
diagrams. Of course, in this way
the ChPT low energy expansion could only be recovered at leading order.
Concerning the divergences, they were regularized with a cutoff, which
violates chiral symmetry, making them finite, but not cutoff independent.
Fortunately, the cutoff dependence was rather weak and the description
of the data was remarkable for cutoffs of the size of the chiral scale. 
Nevertheless, due to this cutoff regularization,
it was not possible to compare
 the eight parameters of the chiral Lagrangian,
which are supposed to encode the underlying QCD dynamics, 
with those obtained from other low energy processes. That is, it was
not possible to test the compatibility of the chiral
parameters with the values already present in the literature.

Of course, due to the controversial nature, or even 
the doubts about the existence of the scalar states,
it is very important to check that
the poles are not just artifacts of the approximations,
to estimate the uncertainties in their parameters, and
to check their compatibility with other experimental information
regarding ChPT. That was the reason why, in a first step, 
the $K\bar{K}\rightarrow K\bar{K}$ one-loop amplitudes
were calculated in \cite{Guerrero:1998ei}, also unitarizing them
coupled to the $\pi\pi$ states, and reobtaining
the $\sigma$, $f_0$ and $\rho$ poles.
The whole calculation of one-loop meson meson scattering 
has been recently completed with the totally new
$K\eta\rightarrow K\eta, \eta\eta\rightarrow\eta \eta$ and
$K\eta\rightarrow K\pi$ amplitudes \cite{GomezNicola:2001as}.
In addition the other five existing independent amplitudes
have also been recalculated. The reason for repeating
those existing calculations is that, to 
one loop, one could choose to write all
amplitudes in terms of just  $f_\pi$, or 
use all $f_\pi$, $f_K$ and $f_\eta$, 
or any other combination of them that is equivalent up to $O(p^4)$
etc... 
However, when one choice is made for one amplitude, 
the other ones have to be calculated consistently in order 
to keep the coupled channel perturbative unitarity, 
which is needed for the IAM. As commented before,
with these unitarized amplitudes we obtained \cite{GomezNicola:2001as} 
a simultaneous description of meson meson scattering data 
in the resonant region up to 1.2 GeV, but also of the
low energy region, with scattering lengths compatible
with the most recent determinations. 
The fact that the calculation was complete to one loop
and renormalized as in standard ChPT, also allowed us
to show that the resulting set of chiral parameters 
was compatible with previous
determinations in the literature.

The final step is therefore to extend analytically the amplitudes
to the complex plane and search for poles in the second Riemann sheet.
We will provide next a brief account of how we have built our amplitudes,
how the data have been fitted, 
but also our first, preliminary, results for the poles, 
although a more detailed
exposition and the final calculations will be presented somewhere else soon
\cite{prep}.

\section{Chiral Perturbation Theory amplitudes}

The QCD  massless Lagrangian for the light $u,d$ and $s$ quarks
is invariant under  the $SU(3)_L\times SU(3)_R$
chiral symmetry, which rotates the Left (or Right) components
of these quarks among them. There is also an small explicit breaking
due to the small masses of those quarks, but at sufficiently high energies
that effect should be rather small. Nevertheless
the $SU(3)_L\times SU(3)_R$ symmetry is not seen in the physical 
spectrum, but only $SU(3)_{L+R}$ is realized approximately
once the small explicit breaking is taken into account. 
The familiar isospin is nothing but the $SU(2)_{L+R}$ subgroup.
The $SU(3)_{L-R}$ symmetry has to be spontaneously broken,
and indeed, the pions, kaons and etas can be identified as the
associated Goldstone bosons of this breaking. Once more, they are not
massless, due to the small masses of those quarks, but they are much
lighter (and much more stable) than 
other hadrons with their same quantum numbers, and than the 
generic hadronic scale of approximately 1 GeV.

These Goldstone bosons are expected to be the relevant degrees of freedom
at low energies. Their low energy dynamics can then be described 
\cite{weinberg}
by the most general Lagrangian made of pions, kaons and etas, that 
implements the symmetry breaking pattern described above, as well as 
other usual constraints
like Lorentz invariance, locality, etc... This is called
Chiral Perturbation Theory \cite{chpt}, and it corresponds to an expansion
in external momenta, the energy or the mass of the mesons, generically
$p$, over the chiral scale $\Lambda=4 \pi f_\pi\simeq1.2\,$GeV. 
The leading term, $O(p^2)$ is nothing but the  non-linear sigma
model and only depends on the meson masses and the chiral scale $4\pi f$, where
$f$ is the meson decay constant at leading order. 
Since there are no more free parameters, it
is universal, i.e., independent of the detailed mechanism
of symmetry breaking.
It is enough to reproduce the current algebra results of the 60's.
At next to leading order $O(p^4)$, there are
eight terms which now are multiplied by some arbitrary low energy
constants $L_i(\mu)$, also called chiral parameters.
These parameters contain information on the specific dynamics of the
underlying theory, but are also needed for the
renormalization of the divergences that appear at one-loop
when one uses vertices from the lowest order Lagrangian.
This renormalization procedure can be carried out to more loops 
by adding higher order terms in the Lagrangian. In this way
it is possible to obtain finite calculations  order by order,
at the price of including an increasing number of parameters.
However, these new terms will all be suppressed by additional
powers of $p^ 2/\Lambda^2$ so that the lowest orders will
be dominant at low energies. For our purposes it will be enough
to work at one-loop, that is $O(p^4)$, so that we still have
amplitudes with  imaginary parts, as well as the eight $L_i$
parameters that contain information on the specific QCD dynamics.

Therefore, the lowest order, ${\cal O}(p^2)$, meson-meson
scattering amplitudes (called ``low energy theorems'' \cite{weinberg}
because as we have just commented, they only depend on
the symmetry breaking scale) are obtained just
from the tree level diagrams of the lowest order Lagrangian. In
contrast, the calculation of the ${\cal O}(p^4)$ contribution
involves the evaluation of the following Feynman diagrams: 
First, the tree level
graphs with the second order Lagrangian, which depend on the
chiral parameters $L_i$.
Second,
the one-loop diagrams in Fig.1, whose
divergences will be absorbed in the $L_i$
through renormalization.

In particular, those graphs in
Fig.1a provide an imaginary part to ensure perturbative unitarity,
whereas those graphs in Fig.1e, provide the wave function,
mass and \textit{decay constant  renormalizations}.
As we will see the renormalization of the decay
constant will play a subtle role in the determination
of the $f_0(980)$ and  $a_0(980$) pole positions.
Let us then explain this somewhat technical point:
Note that the meson decay constants
$f_\pi\simeq94.4$MeV, $f_K=1.22f_\pi$ and $f_\eta=1.3f_\pi$
only differ at $O(p^4)$ \cite{chpt,GomezNicola:2001as}.  
At leading order, all of them are equal
to the only scale in the Lagrangian, $f$,  which,
after renormalization, is not
directly the physical observable.
As a consequence, if we want to write our amplitudes 
in terms of observable quantities, we could 
substitute $f$ by $f_\pi$ or $f_K$ or $f_\eta$,
or any combination of them. We could even make a different
choice for each amplitude {\it as long as we do not couple
the amplitudes among them}. However, if one
wants to study a coupled channel process,
once a choice is made for one amplitude, the choices
for the coupled amplitudes have to be made consistently, if one wants to
ensure perturbative unitarity.
The same argument would follow for the masses, but they 
already differ at leading order, so that the numerical difference 
is irrelevant compared with the decay constant case.

The one-loop amplitudes of $\pi\pi\rightarrow\pi\pi$ \cite{chpt}, 
$\pi K\rightarrow\pi K$ \cite{Kpi}
and that of $\pi \eta\rightarrow\pi \eta$ \cite{Kpi}
were calculated more than a decade ago, because
the thresholds of these reactions
is low enough to apply the standard ChPT formalism.
As explained in the introduction, the
$K\bar{K}\rightarrow K\bar{K}$ one-loop amplitudes
were calculated in \cite{Guerrero:1998ei}, and
those of
$K\eta\rightarrow K\eta, \eta\eta\rightarrow\eta \eta$ and
$K\eta\rightarrow K\pi$ in \cite{GomezNicola:2001as},
much more recently since their thresholds are much higher 
and they only became interesting when 
the appropriate unitarization methods were developed.
In \cite{GomezNicola:2001as}, the other five 
one-loop amplitudes were recalculated in order to express
all of them in terms of $f_\pi$ only, and ensure
exact perturbative partial wave unitarity, which we explain 
in the next section.

\vspace*{.8cm}
\begin{figure}[htbp]
  \includegraphics[height=.1\textheight]{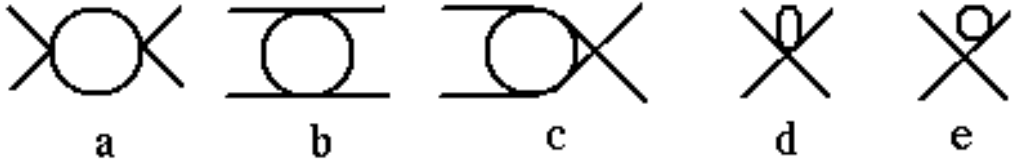}
  \caption{Generic one-loop Feynman diagrams that have to be evaluated
in meson-meson scattering.}
\label{fig1:diagrams}
\end{figure}

As we have already commented in the introduction, 
meson-meson scattering data is customarily presented 
using partial waves
of definite isospin and angular momentum, $t_{IJ}$.
In particular the data is given in terms of the complex phase
of the amplitude, or phase shifts $\delta_{IJ}$
According to our previous discussion,
the meson-meson partial waves within ChPT are thus
obtained as series in the momenta, ( some terms are also multiplied
by chiral logarithms from the loops functions).
Generically, in the
chiral expansion we will then find, omitting the $I,J$ subindices,
$t\simeq t_2+t_4+...$, where $t_2$ and $t_4$ the ${\cal
O}(p^2)$ and ${\cal O}(p^4)$ contributions, respectively.

\section{Partial wave unitarity}

\vspace{-.2cm} The $S$ matrix unitarity relation  $S S^\dagger=1$
translates into simple relations for the elements of the $T$
matrix $t^{\alpha\beta}$ if they are projected into partial waves,
where $\alpha,\beta,...$ denote the different states physically available. For
instance, if there is only one possible state, $\alpha$,  the partial
wave $t^{\alpha\alpha}$ satisfies
\begin{equation}
\ima t^{\alpha\alpha}  = \sigma_\alpha \,\vert \,t^{\alpha\alpha} \vert ^2 \quad
\Rightarrow \quad\ima \frac{1}{t^{\alpha\alpha}}=-\sigma_\alpha\quad\Rightarrow \quad t^{\alpha\alpha} =
\frac{1}{\rea  t^{\alpha\alpha} -i \,\sigma_\alpha}
\label{uni1}
\end{equation}
where $\sigma_\alpha=2 q_\alpha/\sqrt{s}$ and $q_\alpha$ is the C.M. momentum of
the state $\alpha$. Written in this way it can be readily noted that
{\it we only need to know the real part of the Inverse Amplitude}.
The imaginary part is fixed by unitarity. 
As a matter of fact, this
relation {\it only holds above threshold} up to the energy where
another state, $\beta$, is physically accessible.  Above that point,
the unitarity relation for the partial waves can be written
as:
\begin{eqnarray}
\ima t^{\alpha\alpha} &=& \sigma_\alpha \,\vert\, t^{\alpha\alpha} \vert ^2+
\sigma_\beta \,\vert\, t^{\alpha\beta} \vert ^2,\\
\ima t^{\alpha\beta} &=& \sigma_\alpha \, t^{\alpha\alpha}\,  t^{\beta\alpha\;*}+
\sigma_\beta \,t^{\alpha\beta}\, t^{\beta\beta\;*},\nonumber\\
\ima t^{\beta\beta} &=& \sigma_\alpha \,\vert\, t^{\alpha\beta} \vert ^2+
\sigma_\beta \,\vert\, t^{\beta\beta} \vert ^2.\nonumber
\end{eqnarray}
or, in matrix form (and only above the second threshold):
\be
\ima T = T \, \Sigma \, T^* \quad \Rightarrow \quad \ima T^{-1}=- \Sigma
\quad  \Rightarrow \quad T=(\rea T- i \,\Sigma)^{-1}
\label{unimatrix}
\ee
with
\be
T=\left(
\begin{array}{cc}
t^{\alpha\alpha}&t^{\alpha\beta}\\
t^{\alpha\beta}&t^{\beta\beta} \\
\end{array}
\right)
\quad ,\quad
\Sigma=\left(
\begin{array}{cc}
\sigma_\alpha&0\\
0 & \sigma_\beta\\
\end{array}
\right)\,,
\ee
which allows for a straightforward generalization to the
case of $n$ accessible states. Once more, unitarity
means that we would only need to
calculate the real part of the inverse amplitude matrix.

Coming back to ChPT, we can notice that the perturbative 
series of ChPT behave as polynomials with a higher order term 
$O(p^N/\Lambda^N)$. If we substitute them in 
the above unitarity relations for the imaginary parts of $T$, 
which are non-linear,
we will have $O(p^N/\Lambda^N)$ on the left side, but
 $O(p^{2N}/\Lambda^{2N})$ on the right. Hence, ChPT
amplitudes will never satisfy unitarity exactly.
Nevertheless, ChPT partial waves satisfy
unitarity perturbatively, that is, instead of eq.(\ref{uni1}), 
they can satisfy:
\begin{equation}
\ima t_2^{\alpha\alpha}  = 0,\quad
\ima t_4^{\alpha\alpha}  = \sigma_\alpha \,\vert \,t_2^{\alpha\alpha} \vert ^2
\label{pertunit1}
\end{equation}
for the single channel case, 
and  instead of eq.(\ref{unimatrix}), they can satisfy
\begin{equation}
\ima T_2 = 0,\quad \ima T_4 = T_2 \, \Sigma \, T_2^*
\label{pertunitmatrix}
\end{equation}
for the coupled channel case.
Note that , as we did for a single channel,
we are using $T_2$ and $T_4$ for the $O(p^2)$ and 
 $O(p^4)$ contributions to the scattering matrix.
We say ``can satisfy'' because,
generically, the above expressions for the one-loop
contributions do not hold exactly, 
but only up to $O(p^ 6)$.
However,
when expressed in terms of physical decay constants,
the above relations can even be satisfied exactly if 
the substitution of $1/f$ 
in terms of $1/f_\pi$ or $1/f_K$ or $1/f_\eta$
is made to match their corresponding powers on both
sides of the above equations. 
In such case, the $O(p^6)$ can be made to vanish.
(As we already commented, the masses also suffer
the same subtlety and the same care has to be taken with them.)

Since in the literature the amplitudes had been calculated
sometimes just in terms of $1/f_\pi$ but some other times
using or $1/f_K$ or $1/f_\eta$
independently, we recalculated all of them in terms
of just $f_\pi$ in \cite{GomezNicola:2001as}, the simplest choice.
Nevertheless, we are also
presenting here results with the much more natural
choice of using
the decay constants associated to each field in the process.
From the formal point of view, the two choices are equivalent up
to $O(p^4)$, but in the second one the resummation of the decay
constants is implicitly carried out to higher orders.
In addition, it has the advantage of using $f_K$ when dealing with kaons
or $f_\eta$ when dealing with etas. Numerically, the differences
could be sizable at high energies when using the unitarized amplitudes.

\section{Unitarization: The Inverse Amplitude Method}

Unitarity is a very important feature of scattering,
and it is even more relevant when dealing with resonances,
which generically saturate the unitarity bounds.
This can be illustrated in the single channel
case, where eq.(\ref{uni1}) implies 
the following unitarity bound: 
$\vert t_{\alpha\alpha}\vert \le1/\sigma_\alpha$.
Moreover, if we sit on top of a BW resonance, at $s=M_R^2$, we see from 
eq.(\ref{BWform}), that the amplitude becomes purely imaginary,
that is Im$\,t_{\alpha\alpha}=\vert t_{\alpha\alpha}\vert$, and therefore, 
in this case eq.(\ref{uni1})  implies $\vert t_{\alpha\alpha}\vert =1/\sigma_\alpha$.
The unitarity bound is saturated. 
Once more, the ChPT amplitudes
if extrapolated to high enough energies, will violate
also this bound, since they behave as polynomials in $s$.

In order to unitarize the ChPT amplitudes 
one of the simplest methods is to
introduce the $\rea T$ in eq.(\ref{unimatrix}), 
calculated as a ChPT expansion
\begin{eqnarray}
  T^{-1}&\simeq& T_2^{-1}(1-T_4 T_2^{-1}+...),\\
\rea  T^{-1} &\simeq&  T_2^{-1}(1-(\rea T_4) T_2^{-1}+...).
\end{eqnarray}
Taking into account the perturbative unitarity conditions,
 eq.(\ref{pertunitmatrix}),
we thus find
\begin{equation}
 T^{IAM}\simeq T_2 (T_2-T_4)^{-1} T_2,
\label{IAM}
\end{equation}
which is the coupled channel Inverse Amplitude Method, which we have 
indeed
used to unitarize {\it simultaneously} the 
whole set of one-loop ChPT meson-meson
scattering amplitudes. Let us remark that
if we reexpand eq.(\ref{IAM}) at low energies,
we recover the vary same chiral expansion, $T^{IAM}=T_2+T_4+...$,
which ensures that we are respecting the QCD chiral
symmetry breaking pattern at low energies.
In addition, it can be easily checked that $T^{IAM}$ satisfies the
partial wave unitarity conditions, eq.(\ref{unimatrix}), {\it exactly},
above the thresholds of all the physically accessible channels.
Let us also mention that the IAM can be also generalized to higher orders 
\cite{Dobado:1996ps,Nieves:2001de}, including the case when the leading
order $t_2$ vanishes \cite{Dobado:2001rv}.

Let us finally remark that the IAM violates crossing symmetry, since
obviously we are treating the right and the left cuts differently.
The largest influence of the worse left cut approximation
is on the closest point to the left cut, that is, the thresholds.
We will see that the IAM threshold parameters are in good agreement
both with data and with standard ChPT 
(which certainly respects crossing symmetry), therefore
the crossing symmetry violation coming from the IAM itself seems
to be small.
However, as we have already explained, the meson-meson data is obtained
using strong extrapolations. Hence, even the data carries its
own amount of crossing violation if errors are not taken into account.
When considering not only threshold data, but also experimental
information in other regions, {\it including their uncertainties}
it can be shown that the IAM yields indeed just 
an small crossing symmetry violation \cite{Nieves:2001de}.

\section{THE INVERSE AMPLITUDE METHOD FIT TO THE SCATTERING DATA}

Once we had all the amplitudes calculated within the
standard ChPT renormalization scheme
(dimensional regularization in the $\overline{MS}-1$ scheme),
we first looked  at the results
using the IAM with previous determinations of the
chiral parameters from other processes (see the ChPT column in Table 1). 
Due to their large error
bars, the uncertainties thus
obtained were rather large, but all the resonant behavior in meson-meson
scattering was clearly recovered.
For the detailed plots, we refer
the reader to \cite{GomezNicola:2001as}, but this already 
suggests that a description of the resonances is possible within
the uncertainty limits of the chiral parameters.

\begin{table}[hbpt]
\begin{tabular}{|c||c|c||c|c|c|}
\hline
  \tablehead{1}{r}{b}{Parameter} &
  \tablehead{1}{c}{b}{
$K_{l4}$ decays} &
  \tablehead{1}{c}{b}{ChPT} &
  \tablehead{1}{c}{b}{IAM I}&  
  \tablehead{1}{c}{b}{IAM II} &
  \tablehead{1}{c}{b}{IAM III} 
\\
\hline
$L_1^r(M_\rho)$
& $0.46$
& $0.4\pm0.3$
& $0.56\pm0.10$ 
& $0.59\pm0.08$
& $0.60\pm0.09$
\\
$L_2^r(M_\rho)$
& $1.49$
& $1.35\pm0.3$ 
& $1.21\pm0.10$ 
& $1.18\pm0.10$
& $1.22\pm0.08$\\
$L_3 $  &
 $-3.18$ &
 $-3.5\pm1.1$&
$-2.79\pm0.14$ 
&$-2.93\pm0.10$
& $-3.02\pm0.06$
\\
$L_4^r(M_\rho)$
& 0 (fixed)
& $-0.3\pm0.5$& $-0.36\pm0.17$ 
& $0.2\pm0.004$
& 0 (fixed)\\
$L_5^r(M_\rho)$
& $1.46$
& $1.4\pm0.5$& $1.4\pm0.5$ 
& $1.8\pm0.08$
& $1.9\pm0.03$
\\
$L_6^r(M_\rho)$
& 0 (fixed)
& $-0.2\pm0.3$& $0.07\pm0.08$ 
&$0\pm0.5$
&$-0.07\pm0.20$\\
$L_7 $  & $-0.49$ & 
$-0.4\pm0.2$&
$-0.44\pm0.15$ &
$-0.12\pm0.16$&
$-0.25\pm0.18$
\\
$L_8^r(M_\rho)$
& $1.00$
& $0.9\pm0.3$& $0.78\pm0.18$ 
&$0.78\pm0.7$
&$0.84\pm0.23$\\
\hline
\end{tabular}
\caption{Different sets of chiral parameters ($\times10^{3}$).
The first column comes from recent analysis of $K_{l4}$ decays
\cite{BijnensKl4} ($L_4$ and $L_6$ are set to
zero). In the ChPT column $L_1,L_2,L_3$ come from
\cite{BijnensGasser} and  the rest from  \cite{weinberg}. The 
three last 
ones correspond to the values from the IAM including the
uncertainty due to different systematic error used on different
fits. Sets II and II are obtained using amplitudes 
expressed in terms of $f_\pi$, $f_K$ and $f_\eta$, whereas
the amplitudes in set I are expressed in terms of $f_\pi$ only.} \label{eleschpt}
\end{table}

Of course, a much better description could be obtained 
with a fit to the data. We
therefore carried out a fit, using MINUIT \cite{MINUIT},
to the presently available data on meson-meson scattering.  
Due to the already commented problems with the systematic uncertainties
in the data, which has not been quantified in the original articles,
we performed fits adding 
a $1\%$, $3\%$ or a $5\%$ systematic error.
The resulting curves 
are basically indistinguishable to the naked eye.
The errors quoted in 
Table 1 for the IAM sets of fitted chiral parameters, 
correspond to those of MINUIT combined with 
a systematic error
that covers the spread of values obtained
when adding that $1\%$, $3\%$ or $5\%$ systematic error. 
Note that the values we obtain  are compatible 
with previous determinations. 
In particular, we show in 
Table 2 the threshold parameters 
compared with existing data and plain ChPT determinations
to one and two loops.

\begin{table}[h]
\begin{tabular}{|c|c|c|c|c|}
\hline 
Threshold&Experiment&IAM fit I&ChPT ${\cal O}(p^4)$&ChPT ${\cal O}(p^6)$\\
parameter&&\cite{GomezNicola:2001as}&\cite{Dobado:1992ha,Kpi}
&\cite{Amoros:2000mc}\\ 
\hline \hline 
$a_{0\,0}$&0.26 $\pm$0.05&0.231$^{+0.003}_{-0.006}$&0.20&0.219$\pm$0.005\\
$b_{0\,0}$&0.25 $\pm$0.03&0.30$\pm$ 0.01&0.26&0.279$\pm$0.011\\
$a_{2\,0}$&-0.028$\pm$0.012&-0.0411$^{+0.0009}_{-0.001}$&-0.042&-0.042$\pm$0.01\\
$b_{2\,0}$&-0.082$\pm$0.008&-0.074$\pm$0.001&-0.070&-0.0756$\pm$0.0021\\
$a_{1\,1}$&0.038$\pm$0.002&0.0377$\pm$0.0007&0.037&0.0378$\pm$0.0021\\ 
$a_{1/2\,0}$&0.13...0.24&0.11$^{+0.06}_{-0.09}$&0.17&\\
$a_{3/2\,0}$&-0.13...-0.05&-0.049$^{+0.002}_{-0.003}$&-0.5&\\
$a_{1/2\,1}$&0.017...0.018&0.016$\pm$0.002&0.014&\\
$a_{1\,0}$&&0.15$^{+0.07}_{-0.11}$&0.0072&\\ 
\hline
\end{tabular}
\caption{ Scattering lengths $a_{I\,J}$ and slope parameters
$b_{I\,J}$ for different meson-meson scattering channels. For
experimental references see \cite{GomezNicola:2001as}. Let us
remark that our one-loop IAM results are very similar
to those of two-loop ChPT.}\label{elesfit}
\end{table}

The IAM I fit was obtained 
expressing all the amplitudes in terms of just $f_\pi$,
which, as we have already explained is somewhat unnatural 
when dealing with kaons or etas.  
The plots and the uncertainties of this fit
were already given in \cite{GomezNicola:2001as}, and therefore
we have preferred to present
here our first results using
amplitudes written in terms of 
$f_K$ and $f_\eta$ when dealing with processes
involving kaons or etas.
In particular, we have rewritten
our $O(p^2)$ amplitudes changing one factor of $1/f_\pi$ by
$1/f_K$ for each two kaons present
between the initial or final state, or by $1/f_\eta$ 
for each two etas appearing between the initial and final states.
In the special case $K\eta\rightarrow K\pi$ we have changed 
$1/f_\pi^2$ by $1/(f_Kf_\eta)$.
Of course, these changes introduce some corrections at $O(p^4)$
which can be easily obtained using the relations between
the decay constants and $f$ provided in \cite{chpt,GomezNicola:2001as}. 
The $1/f_\pi$ factor in each loop function  at $O(p^4)$
(generically, the $J(s)$
given in the appendix of \cite{GomezNicola:2001as})
have to be changed according to eqs.(\ref{pertunitmatrix}).
The amplitudes thus obtained are formally equivalent
to the previous ones, up to $O(p^6)$ differences.
However, at high energies there can be some
small numerical differences when determining the poles.
Obviously, the $\pi\pi\rightarrow\pi\pi$ amplitude remains unchanged.

The  fit results using these more naturally normalized amplitudes
are given in Fig.2, and the resulting new sets of parameters is 
also presented in Table 1 as the IAM set II. 
Note that the only parameters that
suffer a sizable change are those related to the definition of
decay constants: $L_4$ and $L_5$. As it happened
in \cite{GomezNicola:2001as}, the uncertainty bands
are calculated from a MonteCarlo Gaussian sampling (1000 points)
of the $L_i$ sets within their error bars, assuming they are
uncorrelated (and therefore they are conservative estimates).

\begin{figure}[hbpt]
\includegraphics[height=0.85\textheight]{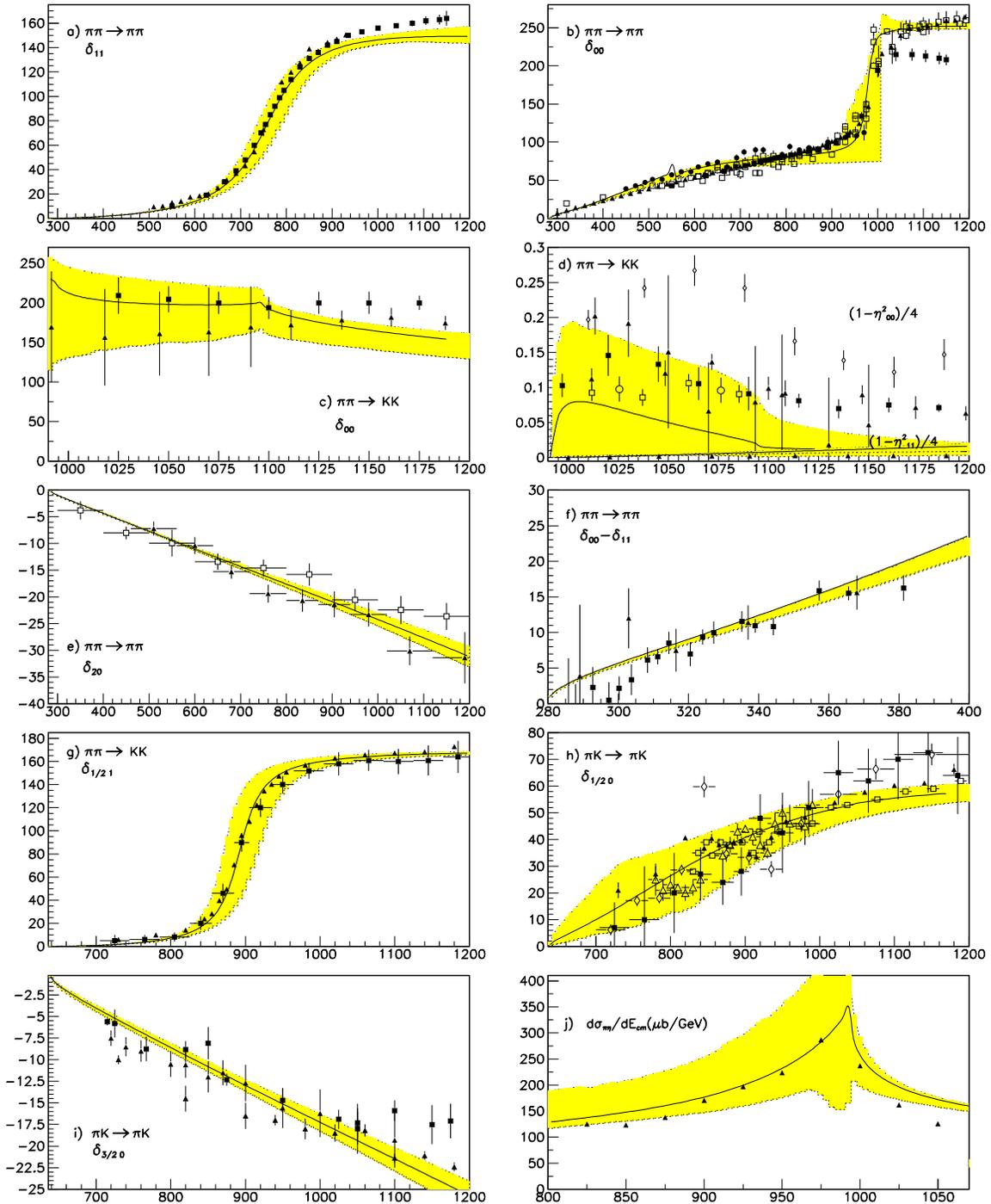}
\caption
{ IAM fit to meson-meson scattering data, set II.
The uncertainties cover also the estimated systematic errors.
The statistical errors from the fit would be much smaller.} 
\label{fig:Tps}
\end{figure}

We have even performed a third fit, the IAM III, by fixing $L_4$ to zero as in 
the most recent $K_{l4}$  $O(p^4)$ determinations given also in Table 1.

Let us recall that in these proceedings we are still showing some 
preliminary results whose calculation is still in progress
\cite{prep}. In a forthcoming work \cite{prep}
we will provide the final numbers (mostly
for the errors) and
the threshold parameters for these other fits. 
Concerning the threshold parameters
we do not expect relevant changes compared to data
 since the $\pi\pi\rightarrow\pi\pi$
amplitude has not changed and therefore the new numbers will remain almost
identical to those of IAM I.

As we can see in Fig.2, we obtain again a nice description 
of meson-meson data up to 1.2 GeV, including once more all the resonant 
behaviors.  One may wonder what would be the effect of applying the
IAM to higher orders. Only the $\pi\pi\rightarrow\pi\pi$ amplitude
has been calculated up to $O(p^6)$ and it has been unitarized
in \cite{Nieves:2001de}, using the higher order form of IAM.
The results regarding poles and resonances in the single channel case
are unchanged and the parameters are compatible
with those of standard ChPT at $O(p^6)$.

Finally, let us remark that the IAM has also 
been applied to $\pi\pi$ elastic scattering in the $(I,J)=(0,2)$ wave
\cite{Dobado:2001rv}, 
whose leading order vanishes.
The amplitude has to be considered up to $O(p^6)$ and  
add an approximation at $O(p^8)$, but the IAM is able to 
generate a pole associated to the $f_2(1200)$ BW resonance.
The mass and widths are in fairly good agreement with data
taking into account that that resonance has only an 80\%
decay into pions.

\section{Poles in meson-meson scattering}

In Table 3 we present the position of poles in the second
Riemann sheet of meson-meson scattering calculated with the 
one-loop IAM. The names we provide
refer to the most similar states that we have found in the literature,
but that does not mean that from the present approach
we could drag any conclusion on their nature.
In Table 4 we provide either the mass and width of these resonances
or  their pole position as given in the PDG.

\begin{table}[htbp]
\begin{tabular}{ccccccc}
\hline
$\sqrt{s_{pole}}$(MeV)
&$\rho$
&$K^*$
&$\sigma$
&$f_0$
&$a_0$
&$\kappa$
\\ \hline
$^{\hbox{IAM Approx}}_{\;\;\;
\hbox{(no errors)}}$
&759-i\,71
&892-i\,21
&442-i\,227
&994-i\,14
&1055-i\,21
&770-i\,250
\\\hline
IAM I
&760-i\,82
&886-i\,21
&443-i\,217
&988-i\,4
& cusp?
&750-i\,226
\\
(errors)
&$\pm$ 52$\pm$ i\,25
&$\pm$ 50$\pm$ i\,8
&$\pm$ 17$\pm$ i\,12
&$\pm$ 19$\pm$ i\,3
&
&$\pm$18$\pm$i\,11
\\ \hline
IAM II
&754-i\,74
&889-i\,24
&440-i\,212
&973-i\,11
&1117-i\,12
&753-i\,235
\\
(errors)
&$\pm$ 18$\pm$ i\,10
&$\pm$ 13$\pm$ i\,4
&$\pm$ 8$\pm$ i\,15
&$^{+39}_{-127}$ $^{+i\,189}_{-i\,11}$
&$^{+24}_{-320}$ $^{+i\,43}_{-i\,12}$
&$\pm$ 52$\pm$ i\,33\\\hline
IAM III
&748-i68
&889-i23
&440-i216
&972-i8
&1091-i52
&754-i230
\\
(errors)
&$\pm$ 31$\pm$ i\,29
&$\pm$ 22$\pm$ i\,8
&$\pm$ 7$\pm$ i\,18
&$^{+21}_{-56}$$\pm$ i\,7
&$^{+19}_{-45}$ $^{+i\,21}_{-i\,40}$
&$\pm$ 22$\pm$ i\,27\\\hline
\end{tabular}
\caption{ Pole positions (with errors) in meson-meson scattering.
When close to the real axis the mass and width of the 
associated resonance is $\sqrt{s_{pole}}\simeq M-i \Gamma/2$.}
\end{table}

\begin{table}[htbp]
\begin{tabular}{ccccccc}
\hline
PDG2002
&$\rho(770)$
&$K^*(892)^\pm$
&$\sigma$ or $f_0(600)$
&$f_0(980)$
&$a_0(980)$
&$\kappa$
\\ \hline
Mass (MeV)
&$771\pm0.7$
&$891.66\pm0.26$
&(400-1200)-i\,(300-500)
&$980\pm10$
&$980\pm10$
&not\\
Width (MeV)
& $149\pm0.9$
& $50.8\pm0.9$
&(we list the pole)
&40-100
&50-100
&listed\\\hline
\end{tabular}
\caption{ Mass and widths or pole positions 
of the light resonances quoted in the PDG.
Recall that for narrow resonances $\sqrt{s_{pole}}\simeq M-i \Gamma/2$}
\end{table}

Let us briefly comment Table 3. In the first line we are giving
the results already obtained in \cite{Oller:1997ng}, with
the approximated coupled channel IAM, using amplitudes
with $f_\pi$, $f_K$ and $f_\eta$. It can be noticed that there were
nine scalar poles, the $\sigma$, the $f_0(980)$, 
the three states of the $a_0(980)$ as well as the four states of
the $\kappa$.  Since they were generated simultaneously, they could be
a good candidate for a nonet, although clearly some mechanism
should be producing the mass difference, very likely some kind of mixing
with higher order states \cite{Oller:1998zr}.

Concerning the results of the IAM, we see that there are always
poles associated to the vector resonances $\rho$ and $K^*$,
in good agreement with the data and with the approximated method.
The uncertainties in the pole positions have been obtained 
again using a MonteCarlo Gaussian sample (300 samples)
of the $L_i$ parameters, within the errors of each set.
Let us note that the vector octet is complete,
since we also obtain a pole in the $(I,J)=(0,1)$
below the $\bar{K}K$ threshold, 
but it is only a crude approximation to the $\Phi$  and $\omega$ states
(it is the octet $\Phi$ indeed). The problem here is that the other relevant
coupled channel that separates the $\Phi$ and the $\omega$
is a three pion state, that we cannot implement in the IAM.
For details, we refer the reader to 
\cite{Oller:1999ag,Oller:1997ng,GomezNicola:2001as}.

Concerning scalar states, from Table 3 we see that the results
concerning the most controversial ones are consistent
and in very good agreement between different IAM sets and
also with the approximated IAM. In other words, the results for the
$\sigma$ and the $\kappa$ poles are robust within this approach:
\textit{there are always ``light'' poles in the $(I,J)=(0,0),(1/2,0)$ channels,
and their  position is fairly well determined},
in round numbers, around $440- i\,215$ MeV for the $\sigma$
and $750-i\,230$ MeV for the $\kappa$. The errors are comparatively
small as it can be seen in Table 3.

The situation concerning $f_0$ is also rather stable for the mass,
which is always around 975 MeV. In contrast, the uncertainty
on the width is rather large. In particular, the central value
is somewhat small when using set 1 (just one $f_\pi$)
but in a fairly good  agreement with data when
considering sets 2 and 3 or the approximated IAM 
(all of them  use $f_\pi,f_K$ and $f_\eta$). As we argued before,
it was natural to expect that 
the use of $f_K$ and $f_\eta$ when dealing with kaons
or etas would provide better results.

Finally, the most sensible state seems to be the $a_0(980)$ resonance.
It can be noticed that it is present as a pole in the second Riemann sheet
 in sets 2 and 3 as well as in the approximated IAM. However, it
is not found as a pole with set 1, using just $f_\pi$.
The fact that the  $a_0(980)$ pole was absent
if one uses only the tree level terms and the tadpoles
of the complete amplitudes in \cite{GomezNicola:2001as}
(again using just $f_\pi$)
with the approximated IAM was first  noted in \cite{Uehara:2002nv}
and has been interpreted as a possible cusp effect.

Given the uncertainty on the $a_0(980)$ it is hard to identify 
it conclusively as a pole or a cusp.
However, we think that there is a somewhat stronger support
for the pole interpretation, although with a strong threshold distortion: On the one hand,
the width of the $f_0(980)$, which is closely related to the $a_0(980)$,
is much better described by the IAM when using several decay constants,
which then give a pole for the $a_0(980)$. On the other hand
the existence of the $a_0(980)$ state seems much less controversial from
other sources apart from meson-scattering data \cite{PDG}.
We remark, anyway, that the two possibilities can be accommodated within 
the IAM.

\section{Conclusions}

We have reported on our recent work where we have completed 
the meson-meson scattering amplitudes to one-loop 
within Chiral Perturbation Theory (ChPT). In order to extend the applicability
of these amplitudes to the resonance region, we have 
unitarized them with the Inverse Amplitude
Method (IAM). In this way, we have been able to describe 
the meson-meson scattering
data up to 1.2 GeV, generating the resonant behaviors, but
simultaneously respecting the chiral low energy expansion.
These new amplitudes are unitarized in dimensional regularization
in order to preserve chiral symmetry, avoiding the use of a cutoff.
Thus we have been able to check 
that the chiral parameters obtained from the IAM
description are compatible with previous determinations 
from other processes within standard ChPT.

In this workshop we have also shown our progress in
determining the position of the poles that appear 
in the IAM amplitudes.
When they are close to the real axis above threshold,
the position of these poles is related to the mass and width
of the associated narrow BW resonances.

In this way, we have been able to establish more robustly
our results for the controversial $\sigma$ and $\kappa$
scalar states. They seem to be generated
simultaneously with the $f_0(980)$ and the $a_0(980)$,
and are therefore good candidates for a possible light scalar nonet.
Nevertheless, the $a_0(980)$ is found to be very sensible to
the choice on how to express the amplitudes
in terms of the physical meson decay constants.

We hope these results
could be of interest in the field of meson spectroscopy


\begin{theacknowledgments}
A.G.N and J.R.P wish to thank the organizers of the "2nd
International  Workshop on Hadron Physics" for their kind invitation
and for their efforts to offer us such a pleasant and lively workshop
in Coimbra.
Work supported by the Spanish CICYT projects, FPA2000-0956,
PB98-0782 and BFM2000-1326. J.R.P. acknowledges support from the
CICYT-INFN collaboration grant 003P 640.15.
  \end{theacknowledgments}


\bibliographystyle{aipproc}   

\begin{thebibliography}{99}



\bibitem{prep}A.~G\'omez Nicola and J.~R.~Pel\'aez, in preparation.

\bibitem{GomezNicola:2001as}
A.~G\'omez Nicola and J.~R.~Pel\'aez,
Phys.\ Rev.\ D {\bf 65} (2002) 054009

\bibitem{newsigma}
R.~Kaminski, L.~Lesniak and J.~P.~Maillet,
Phys.\ Rev.\ D {\bf 50} (1994) 3145.
R.~Delbourgo and M.~D.~Scadron,
Mod.\ Phys.\ Lett.\ A {\bf 10} (1995) 251.
S.~Ishida {\it et al.}, 
Prog.\ Theor.\ Phys.\  {\bf 95} (1996) 745.
M.~Harada, F.~Sannino and J.~Schechter,
Phys.\ Rev.\ D {\bf 54} (1996) 1991
N.~A.~Tornqvist and M.~Roos,
Phys.\ Rev.\ Lett.\  {\bf 76} (1996) 1575.

\bibitem{Dobado:1992ha}
A.~Dobado and J.~R.~Pelaez,
Phys.\ Rev.\ D {\bf 47} (1993) 4883.
Phys.\ Rev.\ D {\bf 56} (1997) 3057.



\bibitem{PDG}  K. Hagiwara {\it et al.}, Phys. Rev. {\bf D 66}, 010001 (2002).

\bibitem{kappa} R.L. Jaffe, \PR{D15} 267 (1977); \PR{D15}, 281 (1977).
E. van Beveren {\it et al.} \ZP{C30}, 615 (1986).
S. Ishida {\it et al}, Prog. Theor. Phys. 98,621 (1997).
D. Black, A. H. Fariborz, F. Sannino, J. Schechter. 
\PR{D58}:054012,1998. 
E.~van Beveren and G.~Rupp,
Eur.\ Phys.\ J.\ C {\bf 22} (2001) 493

\bibitem{Oller:1997ng}
J.~A.~Oller, E.~Oset and J.~R.~Pelaez,
Phys.\ Rev.\ Lett.\  {\bf 80} (1998) 3452;
Phys.\ Rev.\ D {\bf 59} (1999) 074001
[Erratum-ibid.\ D {\bf 60} (1999) 099906].


\bibitem{vanBeveren:2002vw}
E.~van Beveren and G.~Rupp,
arXiv:hep-ph/0201006.


\bibitem{pipidata}
S. D. Protopopescu {\em et al.}, Phys. Rev. {\bf D7}, (1973) 1279;
P. Estabrooks and A.D.Martin, Nucl.Phys.{\bf  B79}, (1974) 301. 
G. Grayer {\em et al.}, Nucl. Phys. {\bf B75}, (1974) 189. 
D. Cohen, Phys. Rev. {\bf D22}, (1980) 2595. 
W. Hoogland {\it et al.}, \NP{B126} (1977) 109.
M. J. Losty {\it et al.}, \NP{B69} (1974) 185.
 R. Mercer {\em et al.}, \NP{B32} (1971) 381.
 P. Estabrooks {\em et al.}, \NP{B133} (1978) 490.
 H. H. Bingham {\em et al.}, Nucl. Phys. {\bf B41} (1972) 1.
 S. L. Baker {\em et al.}, \NP{B99} (1975) 211.
D. Aston {\em et al.}
Nucl. Phys. {\bf B296} (1988) 493.
 D. Linglin {\em et al.}, \NP{B57} (1973) 64 .


\bibitem{Pislak:2001bf}
S.~Pislak {\it et al.}  [BNL-E865 Collaboration],
Phys.\ Rev.\ Lett.\  {\bf 87} (2001) 221801.

\bibitem{charm}E791 Collaboration,\PRL{86},(2001) 770.
C. Gobel for the E791 Collab. hep-ex/0012009.

\bibitem{chpt}
J.~Gasser and H.~Leutwyler,
Annals Phys.\  {\bf 158} (1984) 142.
Nucl.\ Phys.\ B {\bf 250} (1985) 465.

\bibitem{Dobado:1996ps}
T.~N.~Truong,
Phys.\ Rev.\ Lett.\  {\bf 61} (1988) 2526.
\PRL{67}, (1991) 2260;
A. Dobado, M.J.Herrero and T.N. Truong, \PL{B235} (1990) 134.



\bibitem{Guerrero:1998ei}
F.~Guerrero and J.~A.~Oller,
Nucl.\ Phys.\ B {\bf 537} (1999) 459
[Erratum-ibid.\ B {\bf 602} (2001) 641].


\bibitem{weinberg} S. Weinberg, Physica {\bf A96} (1979) 327.

\bibitem{Kpi}
V. Bernard, N. Kaiser, U.G. Mei{\it ss}ner, \PR{D43} (1991) 2757;
\NP{B357} (1991) 129;  \PR{D44} (1991) 3698.

\bibitem{Nieves:2001de}
J.~Nieves, M.~Pavon Valderrama and E.~Ruiz Arriola,
Phys.\ Rev.\ D {\bf 65} (2002) 036002.


\bibitem{Dobado:2001rv}
A.~Dobado and J.~R.~Pelaez,
Phys.\ Rev.\ D {\bf 65} (2002) 077502.

\bibitem{Oller:1999ag}
J.~A.~Oller, E.~Oset and J.~R.~Pel\'aez,
Phys.\ Rev.\ D {\bf 62} (2000) 114017.


\bibitem{MINUIT}  F. James, Minuit Reference Manual D506 (1994).


\bibitem{BijnensKl4} G. Amor\'os, J. Bijnens and P. Talavera,
\NP{B602} (2001) 87.

\bibitem{BijnensGasser} J. Bijnens, G. Colangelo and J. Gasser,
\NP{B427} (1994) 427.


\bibitem{Amoros:2000mc}
G.~Amoros, J.~Bijnens and P.~Talavera,
Nucl.\ Phys.\ B {\bf 585} (2000) 293
[Erratum-ibid.\ B {\bf 598} (2001) 665].



\bibitem{Oller:1998zr}
J.~A.~Oller and E.~Oset,
Phys.\ Rev.\ D {\bf 60} (1999) 074023.





\bibitem{Uehara:2002nv}
M.~Uehara,
arXiv:hep-ph/0204020.
\end{thebibliography}

\end{document}